\documentclass{article}
\usepackage{graphicx} % Required for inserting images
\usepackage{amsmath}
\usepackage{amssymb}
\usepackage{amsthm}
\usepackage{mathrsfs}
\usepackage{epigraph}
\usepackage{url}

\theoremstyle{definition}
\newtheorem{dfn}{Definition}[section]
\newtheorem{prop}[dfn]{Proposition}

\newtheorem{thm}[dfn]{Theorem}
\newtheorem{cor}[dfn]{Corollary}
\newtheorem{rem}[dfn]{Remark}

\newtheorem{axm}[dfn]{Axiom}

\title{Defining the Scope of Learning Analytics: An Axiomatic Approach for Analytic Practice and Measurable Learning Phenomena}
\author{
    Kensuke Takii\thanks{Academic Center for Computing and Media Studies, Kyoto University, Japan} \footnote{kensuke.takii96@gmail.com} \and
    Changhao Liang\footnotemark[1] \and
    Hiroaki Ogata\footnotemark[1]
}
\date{December 2025}

\begin{document}

\maketitle

\begin{abstract}
    Learning Analytics (LA) has rapidly expanded through practical and technological innovation, yet its foundational identity has remained theoretically under-specified. This paper addresses this gap by proposing the first axiomatic theory that formally defines the essential structure, scope, and limitations of LA. Derived from the psychological definition of learning and the methodological requirements of LA, the framework consists of five axioms specifying discrete observation, experience construction, state transition, and inference. From these axioms, we derive a set of theorems and propositions that clarify the epistemological stance of LA, including the inherent unobservability of learner states, the irreducibility of temporal order, constraints on reachable states, and the impossibility of deterministically predicting future learning. We further define LA structure and LA practice as formal objects, demonstrating the sufficiency and necessity of the axioms and showing that diverse LA approaches---such as Bayesian Knowledge Tracing and dashboards---can be uniformly explained within this framework. The theory provides guiding principles for designing analytic methods and interpreting learning data while avoiding naive behaviorism and category errors by establishing an explicit theoretical inference layer between observations and states.
    % This work positions LA as a science of state-transition systems grounded in observability, establishing a foundation for more rigorous theoretical, methodological, and empirical development.
    % ``the science of state transition systems based on observability.''
    This work positions LA as a rigorous science of state transition systems based on observability, establishing the theoretical foundation necessary for the field's maturation as a scholarly discipline.

\end{abstract}

% \epigraph{There is nothing so practical as a good theory.}{--- Kurt Lewin (1890--1947)}

\section{Introduction}
The proliferation of Information and Communication Technology (ICT) in educational settings has led to an explosive increase in educational data, giving rise to a new academic field: Learning Analytics (LA). With the goal of ``understanding and optimizing learning and the environments in which it occurs,'' \cite{ferguson2012learning} LA has created numerous practical success stories: for example, retention prediction \cite{shafiq2022student, li2022retention}, identifying at-risk students \cite{akccapinar2019using}, grade prediction \cite{sghir2023recent}, dashboard construction \cite{larrabee2019efficacy}, categorizing learner behavior patterns \cite{peach2019data}, and building learner models \cite{casalino2021deep, takii2024oklm}. However, behind its remarkable success and development in practice, LA has faced a fundamental problem: it lacked a theoretical foundation \cite{chen2015theory, stewart2017learning, giannakos2023role}.

The lack of a theoretical foundation has led LA to face the following issues. First, it has been unable to clearly define what constitutes LA and what does not among the diverse research on educational understanding and support \cite{rienties2020defining, guzman2021learning, clow2013overview}. As a result, the boundary between what can and cannot be understood through LA has also become ambiguous \cite{gray2022practitioner, selwyn2019s, paolucci2024review}. Such ambiguity leads to a regime that equates the observation of learning behavior derived from data with the learner's characteristics. In other words, it overlooks the essential complexity of the learning process and the fluidity of the learner's internal state \cite{kitto2024places}. This leads to machine behaviorism \cite{knox2020machine}, which equates learning with the control of mechanical actions. Furthermore, there has been little discussion about engaging with learning science, the traditional academic discipline for understanding learning \cite{reimann2016connecting}. LA grew primarily through practice, while its paradigms and academic identity remained ambiguous.

For the reasons stated above, a foundational theory is urgently needed for LA. Specifically, it must be a theory that clearly defines ``what is LA and what is not LA?'' It must be capable of encompassing and supporting diverse LA practices. It should provide justification and explanation for existing LA practices, while offering suggestions for the design and interpretation of future LA practices.

This study provides a theoretical foundation for LA through an axiomatic approach. The axiomatic method has a long tradition in mathematics and theoretical science as a means to clarify the essential structure of a domain \cite{suppes1992axiomatic, hintikka2011axiomatic}. Here, we derive five axioms from the psychological definition of the activity of learning and the various conditions demanded by LA. We also demonstrate that these axioms can fully account for the well-known characteristics of LA and the philosophical demands placed on LA itself. The theoretical framework thus constructed comprehensively explains existing LA and can provide insights for future LA practice.

The contributions of this research are threefold: (1) it explicitly formalizes for the first time the essential properties of LA that were previously known only empirically; (2) it scientifically and philosophically clarifies the scope and limits of observability regarding learning through LA and learner characteristics; and (3) it provides theoretical guidelines for the design of LA practices.

\section{Background}
\subsection{Development and Practices of LA}
According to the definition by SoLAR \cite{WhatisLe37:online} in 2025, LA is:
\begin{quote}
    ``... the collection, analysis, interpretation and communication of data about learners and their learning that provides theoretically relevant and actionable insights to enhance learning and teaching.''
\end{quote}
This definition explicitly states that ``data about learners and their learning'' is the subject of research. Conversely, it implicitly suggests that entities not supplemented by data are not the subject of research. Furthermore, the data targeted is that which yields ``theoretically relevant and actionable insights.'' This clearly indicates that LA is an academic field designed to provide theoretical insights and practical implications. In addition, its purpose is ``to enhance learning and teaching.'' While the expression ``to enhance'' is more ambiguous than ``understanding and optimizing'' in the definition by Ferguson \cite{ferguson2012learning} in 2012, these objectives can be considered to have been carried forward. Overall, LA is a field of study whose fundamental purpose is to contribute to both academic and practical aspects through exploration using learning data, aiming to improve learning and teaching.

As mentioned earlier, LA has produced numerous successful practical applications, but it has not been particularly proactive in establishing a theoretical foundation. This is because LA itself emerged and has been advanced alongside technological progress. The current trends of interest in LA include the following three areas:
\begin{itemize}
    \item \textbf{Interest in Data Science}: The direction of pursuing the potential of data science—specifically, what kinds of analysis can be performed on learning data \cite{baker2014learning, hernandez2022learning}
    \item \textbf{Interest in Learning Improvement}: Practical demands oriented toward learning improvement using results from data analysis \cite{avella2016learning, flanagan2018learning}
    \item \textbf{Interest in Understanding Learning}: An academic directionality that seeks to deepen understanding of the activity of learning based on the results of analysis \cite{winne2017learning}
\end{itemize}
That is, there is a history where the practical aspect of collecting data has been prioritized over creating theory specific to LA. In LA, theory has been treated as something applied to data analysis rather than something created. Indeed, Chen \cite{chen2015theory} argues that LA researchers see themselves as ``relevant adopters of learning theory.''

\subsection{Lack of Theoretical Foundations}
% \begin{itemize}
%     \item では，理論的基盤が不在だと何が起こるのか？ まず，学習に関する伝統的な学習科学との対話が欠如してしまう
%     \begin{itemize}
%         \item LAは本来教育工学的な側面が強い
%         \item 学習に関する立場（行動主義，認知主義，構成主義 \cite{ertmer2013behaviorism}）の特定の立場をとらないで，cherry-picking的に理論を借りてくる
%         \item その結果，学習科学から理論を借りてくることはあっても，LAそのものが理論的貢献をすることは少なかった \cite{dawson2019increasing}
%     \end{itemize}
%     \item また，LAには様々な方法論が乱立し，「どれがLA？」という区別はあいまいになっていた
%     \begin{itemize}
%         \item AIED, CSCL, EDM, LAの分野は相互に重なり合っている \cite{rienties2020defining}
%         \begin{itemize}
%             \item つまり，理論枠組み，方法論，オントロジーが十分に整理されず，境界が不明確になっている
%         \end{itemize}
%         \item 実務的なダッシュボード構築から研究指向のモデリングまで，非常に広い活動がLAとラベル付けされている \cite{hernandez2022learning}
%         \item LAの本質は十分に理論化されてこなかった
%     \end{itemize}
% \end{itemize}
So, what exactly is the substance of the absence of a theoretical foundation in LA? First, dialogue with various academic fields related to learning, especially learning science, is lacking. As mentioned earlier, LA inherently has a strong educational engineering aspect, meaning a strong orientation toward practically understanding and improving learning and education. In both research and practice, it tended not to adopt any specific stance from the various learning perspectives (e.g., behaviorism, cognitivism, constructivism \cite{ertmer2013behaviorism}), instead cherry-picking theories from all fields. Consequently, while it borrowed theories from learning science, LA itself rarely made theoretical contributions \cite{dawson2019increasing}.

Moreover, various methodologies proliferated within the field of LA, and the distinction between what constituted LA and what did not became ambiguous. According to Rienties et al. \cite{rienties2020defining}, the domains of LA, Artificial Intelligence in Education (AIED), Computer Supported Collaborative Learning (CSCL), and Educational Data Mining (EDM) overlap significantly. Theoretical frameworks, methodologies, and ontologies remain insufficiently organized, resulting in unclear boundaries. It has also been pointed out that a very broad range of activities, from practical dashboard construction to research-oriented modeling, are labeled as LA \cite{hernandez2022learning}. The essence of LA has not been sufficiently clarified.

\subsection{Naive Behaviorism}
% \begin{itemize}
%     \item このような理論的基盤の欠如は，学習行動（観測）と学習者の状態（理解・知識）をナイーブに同一視する立場を生むことになった
%     \begin{itemize}
%         \item 行動ログのような観測可能な量から人間の学習を推論するという方法論 \cite{doroudi2024paradigms}
%         \item すなわち，学習の内的プロセスが，データによる学習の捕捉という文脈においては軽視されるのである \cite{knox2020machine}
%         \item 実際はそんなことないのに
%         \begin{itemize}
%             \item 例えば，演習問題に正解する学習行動は，実際にはguessや記憶によっても可能である
%             \item 映像授業の受講は，単に再生しているだけで集中して見ていなくてもデータとして残ってしまう
%             \item 要するに，行動と認知状態は決して一対一対応しないのである！
%         \end{itemize}
%     \end{itemize}
%     \item このような立場は，心理学における古典的行動主義の再来，もしくはデータ至上主義に基づく新しい行動主義と位置付けられる \cite{selwyn2019s}
%     \begin{itemize}
%         \item データからわかる学習者の行動が学習のすべてである
%         \item 学習を行動の変容とみなす素朴な還元主義に堕してしまう
%         \item ゆえに，学習者の行動を機械的に制御すれば，学習は改善できる
%         \begin{itemize}
%             \item これをKnox \cite{knox2020machine} はmachine behaviorismと呼んで痛烈に批判した
%         \end{itemize}
%         \item また，観測できないものは存在しない，という誤謬にも陥ることになる \cite{weidlich2022causal}
%         \item しかし，現代の学習科学は認知的・構成主義的アプローチをも採用している \cite{winne2013nstudy, cakir2022constructivist}
%         \item LAは古典的な行動主義に依拠しているか，データ至上主義に基づいた新たな行動主義に陥っているといえる
%     \end{itemize}
% \end{itemize}

This lack of theoretical grounding led to a position that naively equates learning behavior (observations) with the learner's state (understanding/knowledge). LA essentially adopts a methodology of inferring human learning from observable quantities such as behavioral logs \cite{doroudi2024paradigms}. However, in the context of capturing learning through data, this methodology, if taken too far, leads to the internal processes of learning being neglected \cite{knox2020machine}. In reality, learning is an extremely complex process involving internal mechanisms \cite{kitto2024places}. For instance, the learning behavior of correctly answering practice problems can actually be achieved through rote memorization or recall. Similarly, attending a video lecture---regardless of whether one is concentrating or not---will still be recorded as data if the video is played. In short, learners' actions and cognitive states are never in a one-to-one correspondence.

Thus, the position that equates learners' behavior with their cognitive states can be positioned as a revival of classical behaviorism in psychology, or as a new behaviorism based on data supremacy \cite{selwyn2019s}. In other words, this position holds that the learner's behavior observable from data constitutes the entirety of learning, and therefore learning can be improved by mechanically controlling the learner's behavior. This is a stance that falls into a naive reductionism that views learning as mere behavioral transformation. Knox et al. \cite{knox2020machine} called this ``machine behaviourism'' and subjected it to scathing criticism. Moreover, it leads people into the fallacy that what cannot be observed does not exist \cite{weidlich2022causal}. However, modern learning science also adopts cognitive or constructivist approaches \cite{winne2013nstudy, cakir2022constructivist}. LA is always in danger of falling into machine behaviorism by succumbing to naive behaviorism.

\section{A Theory of Learning Analytics}
\subsection{Requirements for the Axiomatic System}
% \begin{itemize}
%     \item 本研究で設計する公理系は，学習の定義およびLAの哲学的・技術的要請から導かれるものでなければならない
%     \begin{itemize}
%         \item この節では，公理系を設計するにあたって必要となる理論的，および技術的な要請について述べる
%     \end{itemize}
% \end{itemize}
To overcome the current situation described above, this study constructs a formal theory of LA using an axiomatic approach. This study aims to axiomatize LA as the practice of observing and analyzing learning through data, not to design axioms for learning activities themselves. The axiomatic system designed in this study must be derived from the psychological definition of learning, the scientific and technical definition of LA, and the general philosophical requirements for axiom design. This section describes those requirements necessary for designing the axiomatic system. 

\subsubsection{Psychological Definition of Learning}
% \begin{itemize}
%     \item まず，LAが研究対象とする「学習」が，心理学的にどう定義されているのかを見ていく必要がある
%     \begin{itemize}
%         \item 学習とは，一般に「経験によって生じる比較的永続的な行動・知識の変化」と定義されてる
%         \begin{itemize}
%             \item Hilgard and Marquisの伝統的かつ行動主義的定義 \cite{hilgard1940conditioning}
%             \item Mayerの認知主義的定義 \cite{mayer2010applying}
%             \item Schunkの折衷的定義 \cite{schunk1996learning}
%             \item 上記のいずれにおいても，「練習や経験」が「行動やその潜在的可能性，知識」に対して「比較的永続的な変化」をもたらすこととして論じられている
%         \end{itemize}
%     \end{itemize}
%     \item また，学習という過程の中では，経験が原因，内的状態の変化が結果という因果律を持っている
%     \begin{itemize}
%         \item この因果律は時間に巡行する変化であり，時間を逆行して進むことはない
%         \begin{itemize}
%             \item すなわち，行動の変化が経験を生むことは学習とは呼ばれない
%         \end{itemize}
%         \item 以上から，学習を研究対象とするというのは，これらの二つおよびこの因果の機序を対象とするということ
%         \item LAの対象は学習である以上，この因果仮定を扱えなければならない
%     \end{itemize}
% \end{itemize}
First, we need to examine how the activity of learning, which is the subject of LA research, is defined psychologically. Learning is generally defined as ``a relatively permanent change in behavior or knowledge resulting from experience.'' Looking more closely, various definitions actually exist depending on the perspective one adopts (e.g., the traditional and behaviorist definition by Hilgard and Marquis \cite{hilgard1940conditioning}, the cognitivist definition by Mayer \cite{mayer2010applying}, and the eclectic definition by Schunk \cite{schunk1996learning}). However, in each of the above cases, learning is discussed as practice and experience bringing about relatively permanent changes to behavior, its potential, and knowledge.

Furthermore, observing this definition reveals that learning possesses a causal law where experience is the cause and changes in knowledge are the effect. This causal law represents a change progressing forward in time; reversing time---that is, changes in knowledge producing experience---is not called learning. Consequently, making learning the subject of research means focusing on the experience and knowledge that serve as its cause and effect, respectively, as well as the mechanism of this causality.

\subsubsection{Scientific and Technical Requirements by LA}
% \begin{itemize}
%     \item 次に，学習を対象としているLAという研究分野は，科学的ないし技術的に何を要請しているのか？
%     \item まず，LAによる科学的な要請として，データドリブンな方法論に基づくということがある
%     \begin{itemize}
%         \item すなわち，学習データによって観測された学習行動のみを根拠として，学習や教授をenhanceすることがLAという営みである
%         \item 学習行動のデータによる観測から，学習のトリガーとなった経験が構成される
%         \item 学習データは可視化・分析されて，そこから知見が得られる
%         \item データの分析結果は，学習者の内的状態やその時系列を反映しているとする
%         \begin{itemize}
%             \item すなわち，学習者の内的状態は，学習者の真の状態の観察ではなく，データによる観測から構成される構築物として得られるものである
%         \end{itemize}
%         \item また，データ分析によって得られた知見は，``theoretically relevant and actionable''であるようにならなければならない
%         \begin{itemize}
%             \item これは，データの分析結果が反映する学習者の内的状態やその時系列を``theoretically relevant and actionable''な知見とするために，何かしらの評価や価値判断が行われることを意味している
%         \end{itemize}
%         \item したがって，先の学習の定義と合わせると，LAの流れは以下のようになる
%         \begin{itemize}
%             \item 学習行動（観測） → 観測から構成される経験 → 状態 → 評価
%             \item これは図でも説明しておく
%         \end{itemize}
%     \end{itemize}
%     \item LAによる技術的要請について論じる
%     \begin{itemize}
%         \item LAは情報技術を用いたデータ取得を行うことを前提とするため，観測が行われる時間は離散的でなければならない
%         \item LAによる評価の形は限定されてはならない（実数値，整数値，それ以外の定性的な形式など）
%         \item データから観測できない状態については，構成することはできないことが保証される
%         \item 時間的因果の表現のため，データの単なる蓄積のみならず，その系列そのものが本質的な重要性を持たなければならない
%     \end{itemize}
% \end{itemize}
Next, we will examine what the field of LA, which is the subject of our research, requires. First, as a scientific requirement for LA, it must be based on data-driven methodology. That is, LA is the practice of enhancing learning and teaching solely based on learning behaviors observed through learning data. Within the learning mechanism, learning behaviors generate triggering experiences. The resulting learning data, captured from these behaviors, becomes the subject of visualization and analysis, yielding insights. The results of data visualization and analysis are interpreted as reflecting the learner's internal state and its temporal progression. Moreover, insights gained through data visualization and analysis must be ``theoretically relevant and actionable.'' This implies that some form of inference or value judgment is applied to transform the learner's internal state and its temporal sequence, as reflected in the visualization and analysis results, into valuable insights. Therefore, combining this with the earlier definition of learning, the flow of LA is as follows: (1) Learning behavior (observation, $\mathcal{O}$), (2) Experience ($\mathcal{E}$), (3) State ($\mathcal{S}$), and (4) Inference ($\mathcal{I}$) (Table \ref{tab:la_flow}).

\begin{table}[htbp]
    \centering
    \begin{tabular}{lp{250pt}}
        \textbf{LA flow} & \textbf{Description} \\
        \hline
        (1) Observation & Entities that can be observed as learning data \\
                        & (e.g., learning activity logs, test scores, health observation, mental state observation sheet) \\
        (2) Experience  & Meaningful events that trigger learning \\
                        & (e.g., attempting quizzes, mistakes, visiting concepts, checking progress/health/mental dashboards) \\
        (3) State       & Latent, unobservable representation of the learner's status \\
                        & (e.g., knowledge, engagement, health, mental status) \\
        (4) Inference   & Quantities/qualities derived from the learner's state \\
                        & (e.g., proficiency value, visualizations, recommendations, predictions) \\
    \end{tabular}
    \caption{Steps of the flow of LA and their descriptions}
    \label{tab:la_flow}
\end{table}

As technical requirements for LA, first, since LA presupposes data acquisition using ICT, the time at which observations occur must be discrete. Next, the form of inference by LA must not be limited. That is, it must be capable of handling not only integer or real values like test scores, but also qualitative forms such as leveling, classification, recommendation, or prediction. Furthermore, it must be guaranteed that states not observed from the data cannot be constructed. In addition, for the representation of temporal causality, not only the accumulation of data but the sequence itself must possess intrinsic significance.

\subsubsection{Philosophical Requirements}
% \begin{itemize}
%     \item 最後に，公理設計における一般的な哲学的要請を整理する
%     \begin{itemize}
%         \item その公理から得られた理論やモデルが反証可能で，明白な循環を含まないこと
%         \item 単純すぎず，複雑すぎないこと（最小十分性）
%         \item 様々なインスタンスをすべて埋め込めること
%         \begin{itemize}
%             \item ここではLA手法（BKT, ダッシュボードなど）
%             \item すなわち，一般化可能性と実践における実用性を兼ね備えることが必要である
%         \end{itemize}
%         \item 因果律が守られること
%         \begin{itemize}
%             \item ここでは，学習という営み自体の時間的因果律を指している
%         \end{itemize}
%     \end{itemize}
% \end{itemize}
Finally, we organize the general philosophical requirements for axiom design. First, theories and models derived from the axioms must be falsifiable and free from obvious circularity. Second, they must possess minimal sufficiency—neither overly simple nor excessively complex. Third, they must be capable of embedding all possible instances; e.g., methods used in LA. That is, the axioms require both generalizability and practical utility in implementation. Furthermore, temporal causality must be preserved. This refers to the temporal causality inherent in the learning process itself.

\subsection{Axiomatic Foundations and Definitions}
\subsubsection{Axiom System}
\begin{figure}[htbp]
    \centering
    \includegraphics[width=\linewidth]{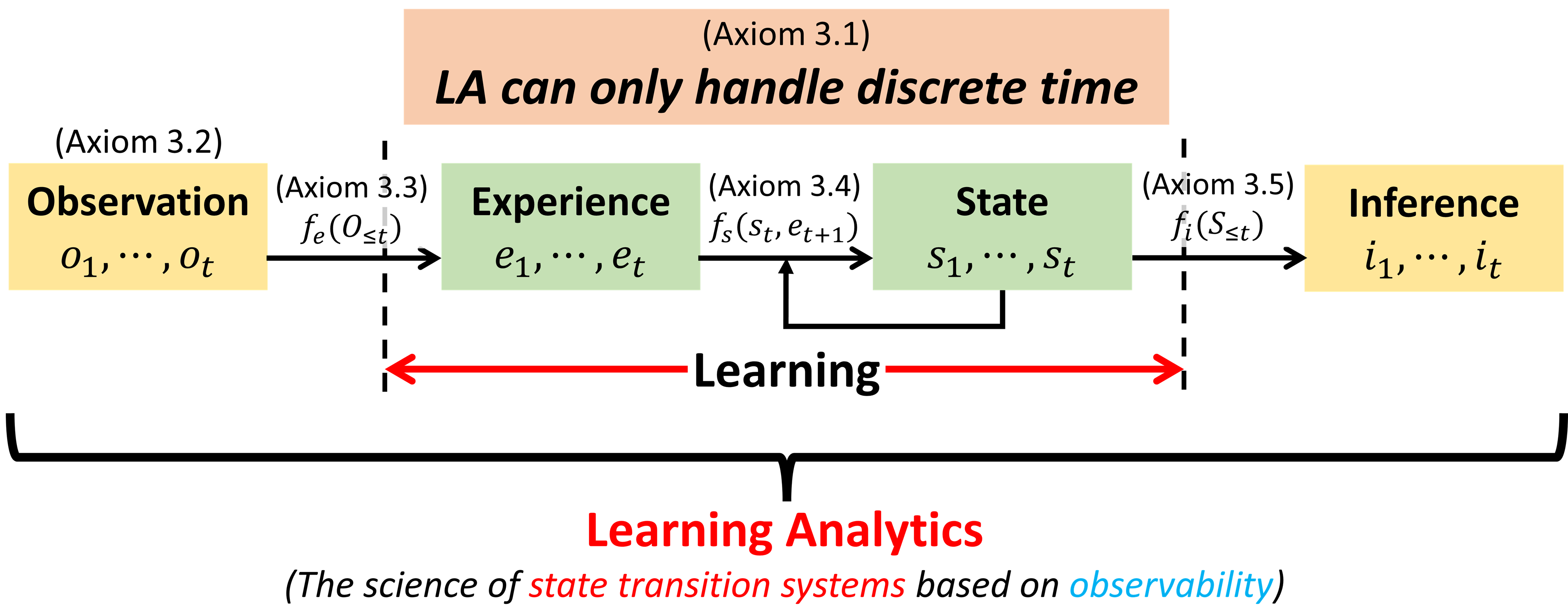}
    \caption{Overall structure of the axiomatic LA, consisting of discreteness of time (Axiom \ref{axm:time}), observation construction (Axiom \ref{axm:observability}), experience construction (Axiom \ref{axm:experience}), state transition (Axiom \ref{axm:state}), and inference (Axiom \ref{axm:inference}).}
    \label{fig:axiom_system}
\end{figure}

Here, we present an axiomatic system designed to satisfy all the above requirements. This axiomatic system contains the following five axioms. Figure \ref{fig:axiom_system} shows the summary of the system.

\begin{axm}[Discreteness of time]\label{axm:time}
    The time observable in LA forms a discrete totally ordered set $\mathbb{T} \subseteq \mathbb{N}_0 = \{0, 1, 2, \ldots\}$ (where $0\in \mathbb{T}$). The order on $\mathbb{T}$ is consistent with the order on $\mathbb{N}_0$.
\end{axm}

\begin{axm}[Observability]\label{axm:observability}
    Regarding a given learner, LA can only handle observations $o_t \in \mathcal{O}$ (where $t \geq 1$) and their sequences $O_{\leq t} = (o_{t_1}, o_{t_2}, \ldots, o_t) \in \mathcal{O}^*$ (where $t_1, t_2 \in \mathbb{T}, t_1 < t_2 < t$ and $\mathcal{O}^*$ is a set of observation sequences). Furthermore, $O_{\leq 0} = \epsilon_o$ (empty sequence of observations). Everything the LA can express must be constructed solely from $O_{\leq t}$.
\end{axm}

\begin{axm}[Experience]\label{axm:experience}
    From an observation sequence $O_{\leq t}$, there exists a mapping $f_e: \mathcal{O}^* \to \mathcal{E}$ that constructs an experience $e_t$. That is, $f_e(O_{\leq t}) = e_t$. Here $f_e$ satisfies the following:
    \begin{itemize}
        \item[(i)]  Determinism: The same observation sequence constructs the same experience
        \item[(ii)] Context dependency: $e_t$ may depend on the entire $O_{\leq t}$ (i.e., it may not be determined by $o_t$ alone)
    \end{itemize}
\end{axm}

\begin{axm}[State and its transition]\label{axm:state}
    A state space $\mathcal{S}$ and an initial state $s_0\in\mathcal{S}$ exist. A state transition map $f_s: \mathcal{S} \times (\mathcal{E} \cup \{ \epsilon_e \})\to \mathcal{S}$ exists, and the state is updated by $s_{t+1}=f_s(s_t,e_{t+1})$. However, $\epsilon_e$ is a symbol that means no experience (See Remark \ref{rem:state_changes}).
\end{axm}

\begin{axm}[Inference]\label{axm:inference}
    For a sequence of states $S_{\leq t}=(s_0,s_1,\cdots,s_t)$, there exists an inference function $f_i: \mathcal{S}^*\to\mathcal{I}$ (where $\mathcal{S}^*$ is a set of state sequences) that constructs an inference value $i_t\in\mathcal{I}$. That is, $i_t = f_i(S_{\leq t})$. Here $f_i$ satisfies the following:
    \begin{itemize}
        \item[(i)]  Determinism: The same state sequence constructs the same inference value
        \item[(ii)] History dependency: $i_t$ may depend on the entire $S_{\leq t}$ (i.e., it may not be determined by $s_t$ alone)
    \end{itemize}
\end{axm}

The difference between experience and state in these axioms lies in whether they are personal or not. That is, experience is constructed purely from data with reproducibility and is unaffected by other factors. However, state can only be inferred from the history of experiences, and even with the same experience, different states lead to different transitions. In other words, even with the same action, its meaning changes depending on the learner's state. This is noted as the following remark.

\begin{rem}[Objectivity of experience vs. Subjectivity of state]\label{rem:exp_vs_state}
    Experience $e_t$ is a non-personal entity deterministically constructed from the observation sequence (Axiom \ref{axm:experience}(ii)). In contrast, state $s_t$ represents the learner's unique internal state, causing different state transitions for the same experience depending on the value of $s_t$. This distinction allows LA to reconcile the objectivity of observations with individual differences in the learning process.
\end{rem}

Furthermore, this axiomatic system assumes state transitions in learners that are not observable in experience. That is, while the specific implementation method is not specified, it inherently includes state changes due to processes such as forgetting or introspection; phenomena occurring reliably within the learner's mind, though unobservable from the outside. This will be demonstrated in the following remark.

\begin{rem}[Expression of state changes not apparent in experience]\label{rem:state_changes}
    Within the domain of the state transition map, the extension of the experience space $\mathcal{E}$ to $\mathcal{E}\cup\{\epsilon_e\}$, which includes the empty experience, represents the reflection of state changes not occurring in experience (e.g., forgetting, introspection). Specifically, this can be expressed as $s_{t+1}=f_s(s_t,\epsilon_e)\neq s_t$.
\end{rem}

\subsubsection{Definitions}
% \begin{itemize}
%     \item 本節では，先の公理系をもとにして，LA構造およびLA実践を定義する
%     \begin{itemize}
%         \item まず，LA構造は，ある学習者の学習の過程（すなわち，経験が内的状態を変化させる過程）に，以下を追加したもの
%         \begin{itemize}
%             \item 学習行動: 学習のトリガーとなる経験をもたらすもので，LAが観測できる唯一の対象
%             \item 評価: 学習の結果変化した内的状態に対して，LAが行う評価．観測されたデータを用いて，何らかの解釈が加わることが一般的
%         \end{itemize}
%         \item これらは，学習者，およびそこで行われているLAの特徴を表現するのに必要十分（後述）
%         \begin{itemize}
%             \item すなわち，観測から経験・状態・評価がどのように構成されるかを決定するのに必要十分である．
%             \item LA構造を，公理系によって定義された，LAを特徴づけるパラメータの組として定義する
%         \end{itemize}
%     \end{itemize}
%     \item また，LA実践では，データの観測から，分析対象となる何らかの評価値を得ることが要請される
%     \begin{itemize}
%         \item さらに，このようにして得られる評価値は一つとは限らない
%         \item 同じデータの系列から様々な分析結果を得るのが一般的である
%         \item これは系列から評価値への単なる写像としてではなく，写像族として定義するのが妥当である
%         \item したがって，データの観測系列から評価値への写像の族として，LA実践を定義する．
%     \end{itemize}
% \end{itemize}

In this section, we define the LA structure and LA practice based on the axiomatic system presented earlier. First, we conceptualize the LA structure as a model that determines how a learner's experience, state, and inference are constructed from observations of their learning. The set of variables defined by the axiom system is necessary and sufficient to construct these features (see Section \ref{subsec:integrity}). Therefore, we define the LA structure as the set of parameters characterizing LA, defined by the axiom system, as follows:

\begin{dfn}[LA structure]\label{dfn:la_structure}
    The nine-tuple defined by
    $$\mathscr{L}=(\mathbb{T}, \mathcal{O}, \mathcal{E}, \mathcal{S}, \mathcal{I}, s_0, f_e, f_s, f_i)$$
    is called an \textbf{LA structure}.
\end{dfn}

From this point onward, to simplify notation without causing misunderstanding, we will refer to $t_1, t_2, \ldots$, which satisfy $t_1, t_2, \ldots \in \mathbb{T}$ and $t_1 < t_2 < \ldots < t$, simply as $1, 2, \ldots$, respectively.

% Finally, the irreducibility of the entire observation sequence in learning is demonstrated. That is, it shows that the order within the observation sequence is an element that cannot be ignored. This suggests that some elements in learning can be lost in LA that ignores the order of the observation sequence.
Here, using the term LA structure, we rewrite Axiom \ref{axm:experience}(ii) (Context dependency) and \ref{axm:inference}(ii) (History dependency) as follows to avoid any ambiguity. These propositions mean the irreducibility of the entire observation or state sequence in learning. That is, they show that the order within the observation or state sequence is an element that cannot be ignored. This suggests that some elements in learning can be lost in LA that ignores the order of the sequence.

\begin{prop}[Context dependency of experiences]\label{prop:exp_ii}
    Axiom \ref{axm:experience}(ii) is equivalent to the following proposition:
    
    ``For a given LA structure
    $$\mathscr{L} = (\mathbb{T}, \mathcal{O}, \mathcal{E}, \mathcal{S}, \mathcal{I}, s_0, f_e, f_s, f_i),$$
    there exists a permutation $\sigma$ that is not the identity mapping, such that for two distinct observation sequences $O_{\leq t}=(o_1, \ldots, o_t), O_{\leq t}^\sigma=(o_{\sigma(1)}, \ldots, o_{\sigma(t)}) \in \mathcal{O}^*$,
    $$f_e(O_{\leq t}) \neq f_e(O_{\leq t}^\sigma)$$
    holds.''
\end{prop}

\begin{prop}[History dependency of inference values]\label{prop:eval_ii}
    Axiom \ref{axm:inference}(ii) is equivalent to the following proposition:

    ``For a given LA structure
    $$\mathscr{L} = (\mathbb{T}, \mathcal{O}, \mathcal{E}, \mathcal{S}, \mathcal{I}, s_0, f_e, f_s, f_i),$$
    there exists a permutation $\sigma$ that is not the identity mapping, such that for two distinct state sequences $S_{\leq t}=(s_1, \ldots, s_t), S_{\leq t}^\sigma=(s_{\sigma(1)}, \ldots, s_{\sigma(t)}) \in \mathcal{S}^*$,
    $$f_i(S_{\leq t}) \neq f_i(S_{\leq t}^\sigma)$$
    holds.''
\end{prop}

Furthermore, in the practice of LA, it is required to derive some inference value from the observed data for analysis. Therefore, it is desirable to adopt a mapping from the observation sequences to inference values as the definition of LA practice. However, the inference values obtained in this manner are not necessarily limited to a single type. Depending on the analytical method, it is common for various types of inference values to be derived from the same data series. Therefore, it is appropriate to view LA practice not as a single mapping from observation sequences to inference values, but as a family of mappings. Consequently, LA practice is defined as follows:

\begin{dfn}[LA practice and LA function]\label{dfn:la_practice}
    % The family of mappings from observation series to inference values, $\mathscr{P}=\{F: \mathcal{O}^*\to\mathcal{I}\}$, is called an \textbf{LA practice}. Functions $F$ that constitute elements of an LA practice are called \textbf{LA functions}.
    Let $\mathscr{P}$ be a non-empty set of functions from $\mathcal{O}^*$ to $\mathcal{I}$:
    $$\mathscr{P} \subseteq \{F\ |\ F: \mathcal{O}^* \to \mathcal{I}\},$$
    which is called an \textbf{LA practice}. Each $F\in \mathscr{P}$ is called an \textbf{LA function}, representing a specific analytic transformation that assigns to every observation sequence $O \in \mathcal{O}^*$ an inference value $i \in \mathcal{I}$.
\end{dfn}

% なお，LA構造には，観測環境によって外生的に決まる要素と，観測対象となる学習者の特徴，および理論の中で設計される内生的構造が両方含まれている．よって，以下のremarkを付け加える．
The LA structure incorporates both exogenous elements determined by the observational environment and endogenous structures designed within the theory, including the characteristics of the learners being observed. Therefore, the following remark is added.

\begin{rem}[External and internal specifications]\label{rem:spec}
    Among the nine components of the LA structure, $\mathbb{T}$, $\mathcal{O}$, $\mathcal{I}$, and $s_0$ represent external specifications (determined by log data, the learner as the observation target, and research design), while the others are internal specifications (elements defining the essence of LA).
\end{rem}

% LAの本質は，上のRemarkにおけるinternal specificationに存在する．すなわち，$\mathcal{E}, \mathcal{S}, f_e, f_s, f_i$はLAが持つ構造を決定するものとして，本質的に高い自由度をもって研究者が決めることができる．したがって，LAの理論的議論は主に$f_e, f_s, f_i$の設計に集中することとなるだろう．
The essence of LA lies in the internal specifications mentioned in the above remark. That is, $\mathcal{E}, \mathcal{S}, f_e, f_s, f_i$ are elements that determine the structure of LA, and researchers can essentially decide them with a high degree of freedom. Therefore, theoretical discussions of LA will primarily focus on the design of $f_e$, $f_s$, and $f_i$.

\subsection{Integrity and Scope of the Axiomatic System}\label{subsec:integrity}
\subsubsection{Sufficiency and Its Proof}
To demonstrate the sufficiency of the given axiom system, we show that a corresponding LA structure can be constructed from any LA practice.

\begin{thm}[Axiomatic system sufficiency]\label{thm:sufficiency}
    Given an arbitrary LA practice $\mathscr{P}\subseteq\{F\ |\ F:\mathcal{O}^*\to\mathcal{I}\}$, there exists an LA structure
    $$\mathscr{L}=(\mathbb{T}, \mathcal{O}, \mathcal{E}, \mathcal{S}, \mathcal{I}, s_0, f_e, f_s, f_i)$$
    satisfying Axioms \ref{axm:time}--\ref{axm:inference}, such that for any $F\in\mathscr{P}$ and any $t\in\mathbb{T}$ and observation sequence $O_{\leq t}$, 
    $$i_t = f_i(S_{\leq t})=F(O_{\leq t})$$
    holds.
\end{thm}
\begin{proof}
    Let $\mathbb{T}$ be a time set, $\mathcal{O}$ an observation space, $\mathcal{I}$ an inference space, and $\mathscr{P}=\{F\}$ an LA practice. Here, $F: \mathcal{O}^*\to\mathcal{I}$ is a deterministic mapping that transforms the observation sequence into inference values. Below, we construct an LA structure satisfying Axioms \ref{axm:time}--\ref{axm:inference}.

    \paragraph{Step 1. Experience Space and Experience Construction}
    Define the experience space as
    $$\mathcal{E}:=\mathcal{O}^*.$$
    Define the experience construction function $f_e:\mathcal{O}^*\to\mathcal{E}$ as the identity function:
    $$f_e(O_{\leq t}):=O_{\leq t}.$$
    By this definition,
    \begin{itemize}
        \item Determinism (Axiom \ref{axm:experience}(i)): The same observation sequence constructs the same experience.
        \item Context dependency (Axiom \ref{axm:experience}(ii)): $e_t=O_{\leq t}$ depends on the entire observation sequence and is not determined by the last observation $o_t$ alone.
    \end{itemize}

    \paragraph{Step 2. State Space and State Transition Construction}
    Define the state space as
    $$\mathcal{S}:=\mathcal{O}^*.$$
    Set the initial state as $s_0 := \epsilon$ (empty sequence). Define the state transition map $f_s: \mathcal{S}\times(\mathcal{E}\cup\{\epsilon_e\})\to\mathcal{S}$ as
    \begin{align*}
        f_s(O_{\leq t-1}, O_{\leq t})&:= O_{\leq t}, \\
        f_s(s, \epsilon_e)&:= s.
    \end{align*}
    By induction, we show that $s_t=O_{\leq t}$ holds for any $t\in\mathbb{T}$.

    \textbf{Base}: When $t=0$, $s_0=\epsilon$ (by definition). 
    
    \textbf{Inductive Step}: Assume $s_{t-1}=O_{\leq t-1}$. By Axiom \ref{axm:experience}, $e_t=f_e(O_{\leq t})=O_{\leq t}$. Therefore,
    $$s_t=f_s(s_{t-1}, e_t) = f_s(O_{\leq t-1}, O_{\leq t})=O_{\leq t}.$$

    \paragraph{Step 3. Inference Function Construction}
    Define the inference function $f_i: \mathcal{S}^*\to\mathcal{I}$ as
    $$f_i(S_{\leq t}):=F(s_t).$$
    Since $s_t=O_{\leq t}$ from Step 2,
    $$f_i(S_{\leq t})=F(O_{\leq t}).$$

    From the construction of each step, it follows that the constructed LA structure
    $$\mathscr{L} = (\mathbb{T}, \mathcal{O}, \mathcal{E}, \mathcal{S}, \mathcal{I}, s_0, f_e, f_s, f_i)$$
    satisfies Axioms \ref{axm:time}--\ref{axm:inference}. Therefore, for an arbitrary LA practice $\mathscr{P}$, an LA structure satisfying Axioms \ref{axm:time}--\ref{axm:inference} can be constructed. Consequently, the axiom system is sufficient for describing LA practices.
\end{proof}

However, it should be noted that the LA structure constructed in this proof is impractical, as it equates experience and state with observation. Here, we have merely formally demonstrated the constructability of the LA structure and, consequently, that any LA practice can be explained within this axiomatic system. Therefore, this system captures the minimal requirements for LA practice, and judgments of whether it is ``good'' or ``bad'' can be made in the derived theorems and principles.

Since such an LA structure can be constructed, the following corollary follows.

\begin{cor}\label{cor:construction_of_la_structure}
    LA functions can be designed to realize a given LA structure. In other words, LA functions $F$ can be realized by a given LA structure $\mathscr{L}$. In this case, we say that $\mathscr{L}$ \textbf{implements the LA function} $F$. In case LA functions implemented by an LA structure $\mathscr{L}$ constitute an LA practice $\mathscr{P}$, we say that $\mathscr{L}$ \textbf{implements the LA practice} $\mathscr{P}$.
    
    Conversely, for any given LA function or LA practice, there exists an LA structure that implements it. However, this structure is not unique; different LA structures may exist that implement the same LA function or LA practice.
\end{cor}

\subsubsection{Necessary Conditions}
% \begin{itemize}
%     \item 以上の公理をすべて満たす任意のLA構造は，以下の性質を必ず持つ．ここではそれらの性質を命題および定理の形で列挙する
%     \begin{itemize}
%         \item 状態の不可観測性 (\ref{prop:unobservability}): 状態は観測そのものではなく，観測の関数である
%         \item 因果的時間構造 (\ref{thm:causality}): 観測は過去の状態を規定するが，未来の状態を決定できない
%         \item 到達可能性制約（\ref{thm:constructible_states}, \ref{cor:unreachable}）: 初期状態から構成可能な状態の集合は，状態空間全体を任意にとることができない
%         \item 経験の順序構造 (\ref{thm:sequence}): 任意の観測列の順序を入れ替えると，一般に経験が変わる
%     \end{itemize}
% \end{itemize}

Next, to demonstrate the necessity of the stated axioms, we describe the properties satisfied by these axioms and how they fulfill the requirements of LA. Any LA structure satisfying all of the above axioms necessarily possesses the properties presented below as theorems. These properties fully express all scientific and technical requirements of LA.

% まず，本公理からは学習者の状態が本質的に直接観測不能で，観測から理論的に構成されたものであることが示される．これが次の命題である．
First, this axiom demonstrates that the learner's state is inherently unobservable and is theoretically constructed from observations. This is the next theorem.

\begin{thm}[Direct unobservability of states]\label{prop:unobservability}
    In the LA framework, the state $s_t$ is not directly observed but is constructively defined from the observation sequence $O_{\leq t}$. That is, $s_t$ itself is not an element of $\mathcal{O}$.
\end{thm}
\begin{proof}
    Axiom \ref{axm:observability} states that LA can handle only observations $o_t\in\mathcal{O}$ and observation sequences $O_{\leq t}\in\mathcal{O}^*$. Axiom \ref{axm:state} states that a state $s_t\in \mathcal{S}$ is constructed from an experience $e_t$. Axiom \ref{axm:experience} states that $e_t=f_e(O_{\leq t})$. Therefore, $s_t$ is a theoretical construct derived from observations and is not a directly observable physical entity.
\end{proof}

% 次に，時間的な因果関係の存在が証明できる．すなわち，観測系列はその時点までの状態を表現することができるが，将来の状態を規定することは不可能である．これは，Predictive LAの実践は現時点までの状態からの外挿であり，未来を確定的に推定することが原理的に不可能であることを示している．
Next, the existence of temporal causality can be demonstrated. That is, while an observation series can represent the state up to that point, it cannot determine future states. This indicates that the practice of predictive LA is an extrapolation from the state up to the present moment, and that it is fundamentally impossible to make deterministic predictions about the future.

\begin{thm}[Causality of time]\label{thm:causality}
    The observation sequence $O_{\leq t}$ can determine the state $s_\tau$ (where $\tau \leq t$) at or prior to time $t$, but it cannot determine the state $s_{\tau'}$ (where $\tau' > t$) after time $t$.
\end{thm}
\begin{proof}
    By Axiom \ref{axm:experience}, the experience $e_t = f_e(O_{\leq t})$ is constructed, and by Axiom \ref{axm:state}, the state $s_t = f_s(s_{t-1}, e_t)$ is constructed. Therefore, $s_t$ is determined by the observation sequence $O_{\leq t}$. Consequently, since the existence of $O_{\leq \tau'}$ is necessary to determine the state $s_{\tau'}$, it is impossible to construct $s_{\tau'}$ solely from $O_{\leq t}$.
\end{proof}

% 次の定理および系は，LAによる観測からでは本質的に構成できない状態が存在することを示す．つまり，LAの原理的な観測限界の存在が示されている．
The following theorem and corollary demonstrate that states exist which cannot be constructed from observations by LA. That is, the existence of fundamental observational limitations for LA is shown.

\begin{thm}[Constraints on constructible states]\label{thm:constructible_states}
    There are structural constraints on the states that can be described within the observable range of LA. Specifically, the state $s_t$ at any time $t$ belongs only to the subset of the state space $\mathcal{S}_\text{reach}(s_0, t)\subseteq\mathcal{S}$ that can be reached from the initial state $s_0$ and the observation sequence $O_{\leq t}$ via the empirical sequence $E_{\leq t}=(e_1, \ldots, e_t)$.
\end{thm}
\begin{proof}
    Define the set of reachable states inductively as follows:
    \begin{align*}
        \mathcal{S}_\text{reach}(s_0, 0) &= \{s_0\}, \\
        \mathcal{S}_\text{reach}(s_0, t+1) &= \{f_s(s, e)\ |\ s \in \mathcal{S}_\text{reach}(s_0, t), e\in\mathcal{E}\cup \{\epsilon_e\}\}.
    \end{align*}

    By Axiom \ref{axm:state}, the state $s_t$ at any time $t$ is represented in the form
    $$s_t = f_s(\cdots f_s(f_s(s_0, e_1), e_2), \cdots, e_t).$$
    By induction, $s_t\in \mathcal{S}_\text{reach}(s_0, t)$. Furthermore, from Axioms \ref{axm:observability} and \ref{axm:experience}, the experience $e_t$ is constructed from the observation sequence $O_{\leq t}$. Therefore, states composed of unobserved experience sequences cannot be described within the LA framework.

    Consequently, for any state $s_t$, $s_t \in \mathcal{S}_\text{reach}(s_0, t)$, and $\mathcal{S}_\text{reach}(s_0, t) \subsetneq \mathcal{S}$ (generally a proper subset).
\end{proof}

\begin{cor}[Existence of an unreachable state]\label{cor:unreachable}
    Generally, in a state space $\mathcal{S}$, there exist states that cannot be reached from any finite time $t$ and any observation sequence. This implies that LA cannot capture the entire potentiality of a learner, but only the realized subset.
\end{cor}

% 最後に，学習における観測系列の全体のessentialityが示される．すなわち，観測系列においてその順番が本質的な要素であることを示している．これは，観測系列の順番を無視したLAでは，学習の本質的な要素が失われてしまっていることを示唆している．
% Finally, the irreducibility of the entire observation sequence in learning is demonstrated. That is, it shows that the order within the observation sequence is an element that cannot be ignored. This suggests that some elements in learning can be lost in LA that ignores the order of the observation sequence.

% \begin{thm}[Irreducibility of temporal order]\label{thm:sequence}
%     The order of observations within an observation sequence is principally irreducible in defining learning experience. Specifically, for a given observation sequence $O_{\leq t} = (o_1, \ldots, o_t)$ and its non-trivial permutation $O_{\leq t}^\sigma = (o_{\sigma(1)}, \ldots o_{\sigma(t)})$ (where $\sigma$ is any permutation other than the identity mapping), $f_e(O_{\leq t})\neq f_e(O_{\leq t}^\sigma).$
% \end{thm}
% \begin{proof}
%     (Proof by contradiction) For all observation sequences, we assume that any arbitrary reordering constitutes the same experience. That is, for any $\sigma$, we assume $f_e(O_{\leq t})=f_e(O_{\leq t}^\sigma)$.

%     At this point, $f_e(O_{\leq t})$ depends only on the set $O_{\leq t}$ of elements of $\{o_1, \ldots, o_t\}$ (multiset) and does not depend on the order. This contradicts Axiom \ref{axm:experience}, which states that $e_t$ may depend on the order structure of $O_{\leq t}$. Therefore, there exist sequences of observations where different arrangements generate different experiences.
% \end{proof}

\subsection{Properties of LA Derived from the Axioms}
\subsubsection{Epistemological Position of LA}
% \begin{itemize}
%     \item 以上の公理系および定理群によって，LAの本質的な特徴や観測可能性，および限界が明らかになった
%     \begin{itemize}
%         \item 続いては，これらに基づいてLAの認識論的な立場について論ずることとする
%     \end{itemize}
%     \item LAは，学習者の内的状態に対して，構成的実在論の立場をとるものとする \cite{wallner2006importance}
%     \begin{itemize}
%         \item すなわち，学習者の内的状態を独立した実在として認めるが，真の実在ではなく，観測可能なものから構成されるものとする立場である
%         \item この点で，LAの立場は素朴実在論 \cite{ross2013naive} とは異なる
%         \begin{itemize}
%             \item 状態を原理的にアクセス不可能ではあるが，実在するものとして認める
%         \end{itemize}
%     \end{itemize}
%     \item しかし，LAの観測可能性にも限界はある
%     \begin{itemize}
%         \item 例えば，未来を予測することはLAの範囲外である
%         \item また，観測によって到達することができる学習者の内的状態も限定的なものであり，データによって表現されない状態について知ることは不可能である
%         \item これは，以下のような実践的含意を含む
%         \begin{itemize}
%             \item LAで「未来が分かる」とするのはそもそも範疇誤謬であること
%             \item Predictive LAにおいて，どんなに予測モデルが高性能になろうとも，やっていることは到達可能な状態集合からの外挿に過ぎず，新しい経験の経路をカバーすることは原理的にできないということ
%         \end{itemize}
%     \end{itemize}
% \end{itemize}
The above axiom system and set of theorems have enabled us to explicitly formalize the essential characteristics, observability, and limitations of LA. Next, we will discuss the epistemological stance of LA based on these foundations.

LA adopts a constructive realist stance \cite{wallner2006importance} regarding the learner's internal state. That is, it is a position that acknowledges the learner's internal state as an independent reality,
% but one that is not a true reality, but rather one constructed from observable phenomena
not as a directly ``true'' reality but as one constructed from observable phenomena
(Theorem \ref{prop:unobservability}). In this respect, the LA position draws a clear line from naive realism \cite{ross2013naive}. In LA, states are fundamentally inaccessible, yet they are recognized as existing entities.

The decisive difference between this research's stance and naive behaviorism lies in the explicitness of the inference process. While naive behaviorism directly equates observation with state, this theory involves the transformation from observation to state via explicit theoretical constructs: the experience-constitution mapping and the state-transition mapping. In other words, this construction process introduces a theoretical inference layer between observations and states, thereby ensuring the verifiability of transformations through the concrete design of mappings and the possibility of multiple state interpretations for the same observation. States are positioned not as observations themselves, but as entities constituted through the theoretical framework used to interpret observations. While the proof of Theorem \ref{thm:sufficiency} formally sets $\mathcal{S}=\mathcal{O}^*$, in practical LA structures, $\mathcal{S}\neq\mathcal{O}^*$, and it is precisely this difference that embodies the distinction between theory and observation.

However, the observability of LA also has its limits. For example, predicting the future falls outside the scope of LA (Theorem \ref{thm:causality}). Furthermore, the learner's internal states that can be reached through observation are also limited (Theorem \ref{thm:constructible_states} and Corollary \ref{cor:unreachable}), and it is impossible to know about states that are not expressed by the data (Axiom \ref{axm:observability}). This has the following practical implications. First, the notion that ``the future can be known'' in LA is a category mistake, or at best, it should be understood as a form of probabilistic inference. Second, in predictive LA \cite{herodotou2019large, sghir2023recent}, which forecasts learners' future states, no matter how sophisticated the predictive model becomes, what is being done is merely extrapolation from the set of reachable states. It is fundamentally incapable of covering pathways of new experiences.

\subsubsection{Principles for LA Practical Design}
% \begin{itemize}
%     \item 次に，これまでに示した公理系および定理群が，LA実践の設計においてもたらす指針を挙げる
%     \begin{itemize}
%         \item これらの指針は単に既存のLA実践に対する説明を与えるだけでなく，将来のLA実践の設計や解釈に示唆を与えることになる
%     \end{itemize}
%     \item 1. 経験，状態，評価はすべて観測のみから構成可能か？
%     \begin{itemize}
%         \item これは公理\ref{axm:observability}から従う
%         \item LAでは観測されたものがすべての前提にならなければならない
%     \end{itemize}
%     \item 2. 経験は観測列全体から構成されているか？
%     \begin{itemize}
%         \item これは公理\ref{axm:experience}(ii)から従う
%         \item 経験がある時点の観測のみに依存するのは不完全なLA実践であり，学習の性質を見逃してしまうかもしれない
%         \item 観測の系列全体から学習経験が構成されなくてはならない
%     \end{itemize}
%     \item 3. 状態遷移は明示的に定義されているか？
%     \begin{itemize}
%         \item これは公理\ref{axm:state}から従う
%         \item 特定の学習者に対しては，同一の観測系列からは同一の状態が構成されなければならない
%         \item これが満たされなければ，データによらないノイズがLA実践に含まれることになってしまい，不適切である
%     \end{itemize}
%     \item 4. 評価は状態の履歴を考慮しているか？
%     \begin{itemize}
%         \item これは公理\ref{axm:inference}から従う
%         \item ここでも，評価はある時点の状態のみならず，状態の時系列全体から構成されなければならない → でないと学習の性質が見逃される
%     \end{itemize}
%     \item 以上の4つの指針は，既存のLA実践が学習の特徴をとらえきれているかどうかを判定するのに役立つだけではなく，将来のLA実践の設計やデータ解釈の手法の厳格化にも寄与することとなるだろう
% \end{itemize}
Next, we present four guiding principles that the axiom system and set of theorems presented thus far provide for the design of LA practices. These principles will serve as recommended conditions that should be met for various practical studies on learning to be considered ``good.'' These guidelines not only offer justification and explanation for existing LA practices but also provide insights for the design and interpretation of future practices.

\paragraph{1. Experience, state, and inference can all be constructed solely from observation}
This is a principle that follows from Axiom \ref{axm:observability}. In LA, everything observed must be a premise.

\paragraph{2. Experience is composed of the entire observation sequences}
This principle follows from Axiom \ref{axm:experience}(ii). Relying solely on observations at a given point in time, especially only the most recent ones, constitutes an incomplete LA practice and risks overlooking the nature of learning. Learning experiences must be constructed from the entire observation sequences, or at least a sufficiently long portion thereof. For example, it is recommended to implement techniques such as setting the width of the sliding window for observations or adopting memory-based models like LSTM instead of simple machine learning models.

\paragraph{3. State transitions are explicitly defined}
This principle follows from Axiom \ref{axm:state}. For any given learner, the same state must be constructed from the same observation sequences. If this is not satisfied, information not derived from the data would be included as noise in the LA practice, which is inappropriate.

\paragraph{4. The inference takes into account the history of the state}
This is a principle that follows from Axiom \ref{axm:inference}. Similarly to Principle 2, inference must be based not only on the state at a single point in time, but on the entire time series of the state, or at least a sufficiently long segment thereof.

These four principles will not only help determine whether existing LA practices adequately capture learning characteristics, but will also aid in designing future LA practices and rigorously refining data interpretation methods.

\section{Case Studies}
% \begin{itemize}
%     \item この章では，ケーススタディとして，実際のLA実践を例示することで，公理系がどのようにLA実践を説明できるのかを論じる
%     \begin{itemize}
%         \item すなわち，この公理系の妥当性を検証するのが目的である
%         \item また，公理を満たさない「悪い」LA実践の例をも提示することで，それらがどのように学習の性質を見逃しているのかについても論じる
%     \end{itemize}
% \end{itemize}
This chapter presents actual LA practices as case studies and discusses how this axiomatic system can explain them. That is, it verifies the validity of this axiomatic system. It also presents examples of naive LA practices that do not satisfy the axioms, discussing how they overlook the nature of learning.

\subsection{Bayesian Knowledge Tracing}
% \begin{itemize}
%     \item BKT \cite{corbett1994knowledge} は，学習者の学習状態をベイズ推定によって隠れ変数を用いて計算する手法である
%     \begin{itemize}
%         \item 学習者が課題に回答するという学習行動をもとにして，課題に対応する知識がどれだけ習得されたのかを推定する
%         \item この推定では，非習得状態から習得状態へ遷移する確率や，当てずっぽうで正解する確率guess，そしてうっかりミスをする確率slipが考慮されている
%         \item BKTは教育工学では広く採用されており，経験的に確立された技術と考えられている
%     \end{itemize}
%     \item では，BKTを本公理系の観点から観察していこう
%     \begin{itemize}
%         \item 観測: BKTによる学習行動の観測は，課題に対するcorrect/incorrect answersに限られる → 観測が時系列的に記録されていく
%         \item 経験: ここでは観測そのものが経験と同一であると言ってよいだろう
%         \begin{itemize}
%             \item correct answer → 課題に関連する知識の習得に寄与した
%             \item incorrect answer → 課題に関連する知識の習得に寄与しなかった
%         \end{itemize}
%         \item 状態: BKTでは，ベイズ推定の隠れパラメータであるlearned/unlearned，および学習者の習熟度を表す習得確率（知識がlearnedである確率）が状態と言える
%         \begin{itemize}
%             \item BKTはパラメータとして初期状態を持っている
%             \item また，経験 (correct/incorrect) から隠れパラメータlearned/unlearnedが確率的に遷移する
%             \item 初期状態から始まって，ベイズ推定によってpre-defined parametersを使いつつ習得確率が更新される
%             \item ベイズ推定であるため，それまでの観測全てがパラメータに反映される
%         \end{itemize}
%         \item 評価: 習熟確率として返す
%         \item なお，時間の取り扱いは，課題への回答時刻を用いているため，離散的である
%     \end{itemize}
%     \item 本モデルでの位置づけ
%     \begin{itemize}
%         \item 以上の観察から，BKTは本公理系を満たすといえる
%         \begin{itemize}
%             \item すなわち，パラメータの初期値さえ設定してしまえば，すべての経験，観測，評価値が観測から構成できる
%         \end{itemize}
%         \item また，観測と経験を同一視している点で，単純なモデルであるともいえる
%     \end{itemize}
% \end{itemize}
The first example is Bayesian Knowledge Tracing (BKT) \cite{corbett1994knowledge}.
BKT is a method that calculates a learner's knowledge acquisition status using Bayesian estimation with hidden variables. It estimates how much knowledge corresponding to a task has been acquired based on the learner's learning behavior of answering the task. This estimation considers initial parameters such as the probability of transitioning from a non-acquisition state to an acquisition state, the probability of guessing correctly, and the probability of making careless mistakes. BKT is widely adopted in educational technology and is considered an empirically established technique.

First, observations of learning behavior in BKT are limited to correct/incorrect answers to tasks. These observations accumulate over time. Therefore, applying this axiom system,
$$\mathcal{O}=\{\mathtt{correct}, \mathtt{incorrect}\}.$$

% Next, regarding experience, it can be equated with the observation itself. That is, a correct answer corresponds to a contribution to acquiring knowledge relevant to the task, while an incorrect answer corresponds to a failure to make such a contribution. Therefore,
Next, an experience can be equated with the observation sequence. In BKT, the knowledge acquisition is probabilistically and chronologically updated by Bayesian estimation using all of the observation sequence. Regarding each observation in the sequence, a correct answer corresponds to a contribution to acquiring knowledge relevant to the task, while an incorrect answer corresponding to a failure to make such a contribution. Therefore,
$$\mathcal{E}=\mathcal{O}^*, \quad f_e(O_{\leq t}) = O_{\leq t}.$$

The BKT state space is a low-dimensional special case of the general state space
$\mathcal{S}$ in TLA, consisting of a binary mastery variable and its associated probability:
$$\mathcal{S}=\{\mathtt{learned}, \mathtt{unlearned}\} \times [0, 1],$$
where the binary mastery represents the latent state of BKT, while probability indicates proficiency.
Thus BKT corresponds to a constrained instantiation of the state-transition structure allowed by the axioms.

Furthermore, the inference space $\mathcal{I}$ in BKT is the interval $[0, 1]$ of possible proficiency levels, which is uniquely determined by the state. That is, the inference value $m_t\in[0,1]$, for example, can be represented as follows:
\begin{align*}
    m_t &= f_i(S_{\leq t}) \\
    &= f_i(((\mathtt{unlearned}, p_0), (\mathtt{learned}, p_1), \ldots, (\mathtt{learned}, p_t))) \\
    &= p_t \in [0, 1].
\end{align*}

Note that the time handled by BKT is discrete, as it uses the time of response to the task.

From the above observations, it is clear that BKT satisfies all axioms of this system. That is, once the initial parameter values are set, all experiences, observations, and inferences are constructed from observations. What is characteristic here is that the sequence of the observations and experience are equated in BKT. From Remark \ref{rem:exp_vs_state}, experience is an impersonal variable independent of individual states. This represents one of the simplest possible designs of the experience space $\mathcal{E}$, where all individual differences are expressed through states (learned/unlearned + probability). This can be considered a valid design satisfying Axiom \ref{axm:experience}.

\subsection{LA Dashboard}
% \begin{itemize}
%     \item 次に，様々な学習ログから取得されたLAダッシュボードを見ていこう
%     \begin{itemize}
%         \item 観測: 多くのLAダッシュボードでは，学習システムへのアクセスログや，クリック履歴，滞在時間などが観測される
%         \begin{itemize}
%             \item すべてログデータから構成される観測量である
%             \item 観測不可能な内的状態を取り扱わない（理解度，モチベーションなど）
%         \end{itemize}
%         \item 経験: 学習システムに依存するが，ログをそのまま使うのではなく，集計，正規化，抽出された特徴量を集める → 観測から経験への写像
%         \item 状態: 内部的には明示的な状態をモデル化しているわけではないが，実際にはすべて累積的指標や時系列的特徴量を通して暗黙の状態を構成している
%         \begin{itemize}
%             \item 例えば，累積学習時間，連続ログイン日数，単元ごとの達成率などから，学習者の理解状態などが暗黙の裡に構成されている
%         \end{itemize}
%         \item 評価: 状態は評価量に射影される
%         \begin{itemize}
%             \item ダッシュボードに表示されるあらゆる値は，学習者に対する評価（理解度の反映，成功体験の反映，パフォーマンスの反映）に対応する
%         \end{itemize}
%         \item なお，ダッシュボードは因果律的順序を反映する
%     \end{itemize}
%     \item このモデルでの位置づけ
%     \begin{itemize}
%         \item 以上の観察より，公理系をすべて満たしている
%         \begin{itemize}
%             \item すなわち，ダッシュボードに表示される内容は，観測から構成されている
%         \end{itemize}
%         \item 特徴として，状態遷移モデルを明示的に定義しないことによって，状態から写像される評価に対する解釈性を広く持たせている
%     \end{itemize}
% \end{itemize}
The next example is an LA dashboard. The visualization of LA dashboard $\mathcal{D}$ can be understood as a set of visualization forms
$$\mathcal{D} \subseteq \mathcal{I}$$
whose elements are human-interpretable visual representations included in an inference space $\mathcal{I}$. This makes dashboards a natural instantiation of the inference axiom.

LA dashboards handle only observations $o_t \in \mathcal{O}$ from log data and their sequences $O_{\leq t} \in\mathcal{O}^*$; unobservable internal states such as comprehension or motivation are never directly handled. Observations acquired in this manner are all obtained at discrete points in time. Furthermore, the design of the observation space $\mathcal{O}$ is system-dependent and is not uniquely determined.

The experience $e_t \in \mathcal{E}$ in the dashboard also depends on the learning system's format, but rather than using the logs directly, they are often stored as aggregated, normalized, and extracted features. In other words, it is a mapping that preserves the temporal structure of the observation. Such aggregations are unique and predefined for each dashboard, and these aggregations are used as experience. This is expressed as a mapping from the observation sequences to the experience:
$$e_t = f_e(O_{\leq t}).$$
At this point, the same experience is constituted within the same observation series, and generally, an experience can depend on the entire observation series.

In LA dashboards, the learner's state $s_t$ is not explicitly modeled internally. However, in practice, states are engineered using cumulative metrics and time-series features based on the learner's first state $s_0$. For example, cumulative learning time, consecutive days logged in, and achievement rates per unit can be considered to construct the learner's understanding state and its transitions. From this perspective, it can be said that there is no LA that does not assume the existence of states.

Moreover, it is clear that the inference is projected onto the state and its transitions. The inference values are then visualized on the dashboard, which can be represented as:
$$\mathcal{D} = \{f_i(S_{\leq t})\ | \ S_{\leq t} \ \mathrm{can \ be\ deduced\ from\ observations}\}.$$

From the above observations, it is evident that the LA dashboards also satisfy all axioms of this system. That is, the content displayed on the dashboard is entirely composed of observations. Furthermore, while it may not explicitly define a state transition model as a feature, this allows users to interpret the inferences mapped from states and their visual representations, thereby enabling broad interpretability.

\subsection{Predictive LA}
The third example is predictive LA, which makes predictions about the learner's future. While predictive LA appears to represent the learner's future state, from the perspective of this theory, it is merely an inference made at a point in time where the future is unobservable. That is, considering Theorem $\ref{thm:causality}$, which demonstrates that future states cannot be determined, the training/testing of predictive LA can be understood as follows. Furthermore, inference refers to any quantity derived from the current state, ranging from simple indicators to complex predictions, recommendations, and optimized decisions.

In predictive LA, the usage logs of the LA system (e.g., system access logs, assignment submissions, behavioral features) constitute the observations $o_t \in \mathcal{O}$, and their sequence $O_{\leq t} \in \mathcal{O}^*$. The observation space $\mathcal{O}$ at this time depends on the system and the type of prediction and cannot be uniquely determined.

Feature generation in predictive LA for model construction is entirely represented as the design of the experience-generating function $f_e$. Specifically, if the feature vector at time $t$ is denoted as $\boldsymbol{x}_t$, then the experience $e_t$ is:
$$e_t = f_e(O_{\leq t}) = \boldsymbol{x}_t.$$
At this point, depending on the prediction model, some utilize only the features at each time step, while others utilize the entire time series (e.g., LSTM, Transformer).

The internal representations of models in predictive LA, such as latent vectors, embeddings, and hidden states of RNNs, can be expressed as elements of the state space $\mathcal{S}$ within the LA structure. That is, they can be represented as, for example, 
$$\mathcal{S} = \mathbb{R}^d. \quad (d\in\mathbb{N})$$
The state transition function $f_s$ can also be expressed as
$$s_{t+1}=f_s(s_t, e_{t+1}) = f_s(s_t, \boldsymbol{x}_{t+1})$$
given the defined state $s_t \in \mathcal{S}$ and the experience $e_{t+1}$ as a feature. In models based on neural or statistical architectures, the state transition function $f_s$ corresponds to the internal update dynamics (e.g., recurrent updates, hidden-state transitions). Furthermore, forgetting or drift during the transition of the learner's states can be incorporated into the design of the function $f_s$.

The predictive results of predictive LA are all expressed by the mapping $f_i$. That is, for the state sequence $S_{\leq t}$ corresponding to the observation sequence, some predictive result (e.g., proficiency level, dropout risk, future performance trajectory) is realized as $i = f_i(S_{\leq t})$.

Here, using the LA structure, training the predictive LA model can be described as optimizing the parametric components of the mappings $f_e$, $f_s$, and $f_i$. Specifically, training involves estimating the feature generation function $f_e$, the function $f_s$ that transitions the model's internal representation, and the prediction output function $f_i$ using past data. Training the predictive LA thus includes the temporal updating of the underlying internal specifications within the LA structure.

Conversely, testing is positioned as verifying how well the LA structure $\mathscr{L}$ constructed through training performs valid inference on unseen observations. Specifically, it applies the obtained observation sequence $O_{\leq t'}$ to the completed mappings $f_e, f_s, f_i$ and outputs the resulting estimate $i' \in \mathcal{I}$. At this point, predictive LA cannot determine the future state itself (Theorem \ref{thm:causality}), but only estimates where the obtained observation corresponds within the reachable states and the inference space $\mathcal{I}$. In this sense, predictive LA does not deterministically predict the future.

From the above observations, predictive LA fully satisfies this axiomatic system. That is, it updates the feature function, internal state, and output function based solely on the observation sequence (training), and checks their consistency against unknown observation sequences (testing). All these elements match the components of the LA structure. Predictive LA cannot determine the future, but it performs probabilistic inference within reachable states and outputs a plausible future.

\subsection{Naive Approaches}
% \begin{itemize}
%     \item この節では，本公理系のうちいくつかの公理を満たさないLA実践の例を挙げ，それがどのように学習の特徴を見落としているのかを論じる
%     \begin{itemize}
%         \item このようなLA実践は，学習データと状態の同一視や，範疇を越えた推論によって，machine behaviorismに陥る危険性をはらんでいる
%     \end{itemize}
%     \item まず，一つ目の例は，単一観測からの状態推定である
%     \begin{itemize}
%         \item 具体的には，学習者の課題に対する回答を，そのまま学習者の理解状態として取り扱ってしまうLA実践である
%         \begin{itemize}
%             \item 課題に正解したら「学習者は理解している」，不正解したら「理解していない」とするもの
%         \end{itemize}
%         \item これは命題\ref{prop:unobservability}に違反している
%         \begin{itemize}
%             \item 状態は観測と同一視されるべきものではなく，観測から構成されるものでなければならない
%             \item このような違反は，データの観測を安易に学習と同一視することで，machine behaviorismへとつながる危険を持っている
%         \end{itemize}
%     \end{itemize}
%     \item 二つ目の例は，順序を無視した分析である
%     \begin{itemize}
%         \item 具体的には，例えば学習者が複数の問題を解いたとき，その正解率を単純な正解数と解答数との比で表すLA実践である
%         \item これは，学習の順序を無視しているという点で，定理\ref{thm:sequence}に違反している
%         \item このような違反では，例えば後半の正解は前半の学習の成果かもしれないが，その情報が失われてしまう
%     \end{itemize}
%     \item 最後の例は，未観測状態への言及である
%     \begin{itemize}
%         \item 例えば，ある学習者がまだシステム上で学習していない分野について，「その分野についてまだ知らない」と推論するLA実践である
%         \item つまり，観測されない部分についての推論をしてしまう
%         \item これは定理\ref{thm:causality}に違反している
%         \begin{itemize}
%             \item 観測されていない状態については，判断できないという立場をとらなければならない
%             \item データとして観測できないところで学んでいるかもしれない
%         \end{itemize}
%     \end{itemize}
% \end{itemize}
Finally, we will discuss examples of ``bad'' LA practices that adopt naive approaches and fail to satisfy several theorems derived from this axiom system, and how they overlook key features of learning. Such practices include equating learning data with states and performing cross-category reasoning.

First, consider an analysis that ignores sequence (Example 1). For example, when a learner solves multiple problems over a long period, this approach represents their accuracy rate as a simple ratio of correct answers to total attempts. This violates Proposition \ref{prop:exp_ii}, which states that the learning sequence must not be ignored. Such a violation means that information is lost: for instance, correct answers in the latter part might be the result of earlier learning, yet this connection is lost.

The next example (Example 2) is state estimation from a single observation. For instance, this involves treating a learner's response to a task directly as an indicator of their understanding state. Specifically, if the learner answers the task correctly, it is considered ``understood''; if incorrect, it is considered ``not understood.'' This violates Theorem \ref{prop:unobservability}, which states that the state is not directly observable. Such a violation leads to the easy conflation of data observation with learning. This carries the risk of falling into machine behaviorism, which posits that learning can be advanced simply by controlling learning behavior. While careless mistakes and guesswork---the kind BKT takes into account---can occur in reality, they are not considered in such a practice.

The final example involves references to unobserved states (Example 3). For instance, an LA practice might infer, ``the learner doesn't know about that field,'' and in some cases, ``therefore, we must support his/her learning in that field,'' regarding a domain a learner has not yet studied on the system. In other words, it is making arbitrary inferences about parts not observed as data. This violates Theorem \ref{thm:causality} concerning temporal causality. LA must adopt the position that it cannot make judgments about unobserved states. LA practices cannot rule out the possibility that learners might be learning in areas not observable as data.

\begin{table}[htbp]
    \centering
    \begin{tabular}{lllllll}
        \textbf{LA practice} & $\mathcal{O}$ & $\mathcal{E}$ & $\mathcal{S}$ & $\mathcal{I}$ & \textbf{Violation} & \textbf{Comment} \\
        \hline
        BKT       & $\checkmark$ & $\checkmark$     & $\checkmark$ & $\checkmark$ & N/A & $\mathcal{E} = \mathcal{O}^*$ (Simplest) \\
        Dashboard & $\checkmark$ & $\checkmark$     & $\checkmark$ & $\checkmark$ & N/A & $\mathcal{S}$: implicit assumption \\
        Predictive LA & $\checkmark$ & $\checkmark$     & $\checkmark$ & $\checkmark$ & N/A & $\mathcal{I}$: Plausible future \\
        Example 1 & $\checkmark$ & $\times$         & $\times$     & $\checkmark$ & Prop. \ref{prop:exp_ii} & Ignores the sequences \\
        Example 2 & $\checkmark$ & $\times$         & $\times$     & $\checkmark$ & Thm. \ref{prop:unobservability} & Equates $\mathcal{O}$ with $\mathcal{S}$ \\
        Example 3 & $\checkmark$ & $\times$         & $\times$     & $\times$     & Thm.  \ref{thm:causality} & Infers unobserved states \\
     \end{tabular}
    \caption{Summary of the case studies.}
    \label{tab:case_study}
\end{table}

Table \ref{tab:case_study} summarizes the results of the case studies. Such case studies serve as corroborating evidence that this axiomatic system can accommodate all LA practices.

\section{Discussion}
% \subsection{学習の構成原理}
% \begin{itemize}
%     \item 本公理系と定理群は，学習を観測されたデータから形式的に扱うとはどういうことかを明確化した
%     \begin{itemize}
%         \item 観測 → 経験 → 状態 → 評価 というLAの構成プロセスの不可避性
%         \item 時間因果律
%         \item 状態の不可観測性
%         \item 順序構造の不可欠性
%         \item 状態空間の到達可能性制限
%     \end{itemize}
%     \item 何がLAで，何がLAでないかに構成的な定義を与えた
% \end{itemize}
% \begin{itemize}
%     \item 以上，これまで本理論における公理系，定理群，およびケーススタディを見てきた
%     \begin{itemize}
%         \item 続いては，このLAの理論が与える貢献や含意，その限界について議論する
%     \end{itemize}
% \end{itemize}
Thus far, we have examined the axiom system, the set of theorems, the implications they suggest, and the case studies within this theory. Next, we will discuss the contributions, implications, and limitations of this LA theory.

\subsection{Principle of Learning Structure}
% \begin{itemize}
%     \item まず，本理論は，学習という複雑な営みを観測されたデータから形式的に扱うとはどういうことなのかを，明示的かつ具体的に形式化した
%     \begin{itemize}
%         \item 「経験が状態を変化させる過程」という学習の定義から始まって，そこに経験をもたらす学習行動の観測，および状態変化に対する評価というLAの方法論を重ねることで，LAを通した学習を形式化したのである
%         \begin{itemize}
%             \item このようなLAの構成プロセスは，原理的に一般的かつ不可避なものである
%         \end{itemize}
%         \item この形式化の結果据えられた5つの公理からは，時間因果律や状態の不可観測性，順序構造の不可欠性と状態空間の到達可能性制限など，LAが直観的かつ暗黙的に共有してきた性質や限界を改めて明確に定式化した
%     \end{itemize}
%     \item そして，LAが明確に公理系として定式化された以上，学習データを伴う様々な研究・実践において，「何がLAで，何がLAでないか」を峻別することが可能になった
%     \begin{itemize}
%         \item すなわち，LA実践が，学習という現象の性質をできる限り見落とさずに捕捉できるようにするための，設計・データ解釈への明確な指針が誕生したのである
%         \item このような指針は，どのようなLA実践が必要十分なのかについて，実践的な示唆を与えるという意味で，単なる概念上の貢献にとどまらない価値を持っている
%     \end{itemize}
% \end{itemize}
First, this theory represents the first explicit and concrete formalization of what it means to formally handle the complex activity of learning from observed data. This formalization begins with the definition of learning as ``the process by which experience changes states.'' By overlaying the LA methodology---observation of learning actions that provide the experience triggering learning, and inference of state changes---we formalized LA itself. Therefore, as a reflection of the definition of learning and the requirements of LA, this LA construction process is, in principle, both general and unavoidable. Furthermore, the five axioms derived from this formalization explicitly restate properties and limitations that LA has intuitively and implicitly shared, such as the indispensability of order structures (Propositions \ref{prop:exp_ii} and \ref{prop:eval_ii}), the unobservability of states (Theorem \ref{prop:unobservability}), temporal causality (Theorem \ref{thm:causality}), and constraints on the reachability of state spaces (Theorem \ref{thm:constructible_states} and Corollary \ref{cor:unreachable}). This theory also clearly distinguishes experience and state based on their attributive nature. This provides a boundary for what can be directly constructed from data and what must be inferred in LA practice.

And now that LA has been clearly formalized as an axiomatic system, it has become possible to sharply distinguish what is LA and what is not in various research and practice involving learning data. In other words, clear principles for design and data interpretation have emerged, enabling LA practice to capture the nature of the learning phenomenon with minimal oversight. These principles hold value beyond mere conceptual contribution in that it provides practical insights into what constitutes necessary and sufficient LA practice.

% \subsection{哲学的含意}
% \begin{itemize}
%     \item 状態は観測されるものではなく構成されるもの
%     \item データ主義LAによって，学習の構成的な側面が見落とされてきた
%     \begin{itemize}
%         \item 観測と状態の同一視の問題
%     \end{itemize}
%     \item LAは本質的に，学習者の真の状態としての内面ではなく，構成された状態としての内面を評価している
%     \begin{itemize}
%         \item LAは学習者の内的状態の存在を否定しない → 観測から構成される別の実態として扱う
%         \item 構成的実在論: 中立的立場
%     \end{itemize}
%     \item LAは本来「log → knowledgeを直接推定する」営みではない
%     \begin{itemize}
%         \item 行動データ自体の否定ではない
%         \item 適切な理論的枠組みがあれば，行動から状態を推論できる
%         \item 本公理系はそのための基盤
%     \end{itemize}
%     \item Machine behaviorismへの反論になっている
% \end{itemize}
\subsection{Philosophical Implications}
% \begin{itemize}
%     \item この理論はLAが学習をどうとらえているのかについても，哲学的な示唆を与える
%     \item まず，LAは本質的に，学習者の真の状態としての内面ではなくて，観測から構成された状態としての内面を評価しているという認識論的立場が，この理論によって明らかになった
%     \begin{itemize}
%         \item 学習者の内的状態を，直接観測できないものとしておきながらも，その存在を一つの実体として否定しないという構成的実在論の立場である．
%         \item 学習データを対象とし，データによる観測のみを推論の根拠とするLAは，しばしば「データからわかることが学習のすべてと誤解している」という批判を受けていた
%         \begin{itemize}
%             \item 観測と状態を同一視してしまうという誤謬に対する恐れがあったためである
%         \end{itemize}
%         \item しかし，本研究によって明らかになったLAの立場は，データおよびそこからわかることを可能な限り誠実に取り扱い，判断可能・不可能な部分を峻別することにつながっている
%         \item その意味で，そのような批判に対する反論として機能している
%     \end{itemize}
%     \item このように，LAはデータから状態を観測することは不可能であるという立場をとるが，逆説的に言えば，適切な理論的枠組みがあれば，行動から状態を「推論」することができることも，本理論によって明らかになっている
%     \begin{itemize}
%         \item すなわち，本理論は既存のLAの価値を否定するものではなく，むしろ理論的根拠を与えることによってenhanceするものである
%         \item 状態の観測不能性は行動データの否定ではない
%         \item 本公理系は，そのように行動から状態を推論し，学習という営為に対する理解や，学習・教育の改善を促進するための理論的基盤となるのである
%     \end{itemize}
% \end{itemize}
This theory also provides a clear answer to the epistemological question of how LA conceives of learning. First, it clarifies the epistemological stance that LA essentially infers the learner's internal state as one constructed from observations, rather than as their true state. This is a position of constructive realism: while treating the learner's internal state as something that cannot be directly observed, it does not deny its existence as a kind of entity. LA, which focuses on learning data and bases its reasoning solely on observations derived from that data, has often faced criticism that it creates the misconception that what can be known from the data constitutes the entirety of learning \cite{selwyn2019s}. However, the LA position clarified by this research leads to treating the data and what can be known from it with the utmost sincerity, while rigorously distinguishing between what can and cannot be judged. In this sense, this theory functions as a response to the criticisms directed at LA.

Thus, LA takes the position that observing learners' internal states from data is impossible. Paradoxically, this also means that with an appropriate theoretical framework, the states can be inferred from actions. That is, this research does not negate the value of existing LA. Rather, by providing theoretical justification for existing LA, it enhances its value. The unobservability of states does not imply the negation of action data. This axiomatic system serves as a theoretical foundation for inferring states from actions, thereby promoting understanding of the learning process and improving learning and teaching.

% \subsection{学習科学との対話}
% \begin{itemize}
%     \item 学習の心理学的定義と整合する
%     \item 行動主義，認知主義，構成主義の立場それぞれの中間をとっている
%     \item 既存の学習理論をLA frameworkでどう表現できるか
% \end{itemize}
\subsection{Dialog with Learning Science}
% \begin{itemize}
%     \item 本理論は，今まで積極的には行われてこなかった学習科学とLAとの対話を促進するためのトリガーともなりうる
%     \item 学習理論は3つのカテゴリに分けることができる \cite{ertmer2013behaviorism}: 行動主義，認知主義，構成主義
%     \begin{itemize}
%         \item 本研究で提案する理論は，これらの3つの立場を統合したものである
%         \begin{itemize}
%             \item 行動主義: データによる学習行動の観測を推論の根拠としている点
%             \item 認知主義: 学習者の状態を観測の系列から構成されるものとして捉え，その変化を取り扱う点
%             \item 構成主義: 観測，経験，状態について，その時系列的変化を本質的要素として取り扱う点
%         \end{itemize}
%         \item その意味で，本理論は相補的ないし対立的だった以上の主義主張をまとめ上げる理論として存在しているのである
%     \end{itemize}
%     \item 本理論は，学習の心理学的定義からスタートした
%     \begin{itemize}
%         \item その意味で，学習に対する既存の科学・心理学的研究と矛盾するものではない
%         \item LAは自らを定式化することで，学習に対する理論的，特に科学的な知見を得るための準備を整えた
%         \item そのため，今までLAが学習科学の知見を借用してきたのと同じように，LAも自らの知見を学習科学に提供できるだけの地盤が整ったのである
%     \end{itemize}
% \end{itemize}
This theory can serve as a trigger to promote dialogue between learning science and LA, which has not been actively pursued until now. According to Ertmer and Newby \cite{ertmer2013behaviorism}, learning theories can be divided into the following three categories: behaviorism, cognitivism, and constructivism. The theory proposed in this study integrates these three positions in the following sense.
\begin{itemize}
    \item \textbf{Behaviorism}: Using data-driven observation of learning behavior as the basis for inference
    \item \textbf{Cognitivism}: Viewing the learner's state as an entity composed of observation sequences and handling its changes
    \item \textbf{Constructivism}: Regarding observations, experiences, and states, treating their temporal changes as irreducible elements
\end{itemize}
In other words, this theory exists as one that synthesizes the three aforementioned positions, which were complementary or even antagonistic.

Furthermore, this theory does not contradict existing scientific and psychological research on learning, as it originates from the psychological definition of learning. By formulating itself, LA has prepared the groundwork for gaining theoretical, particularly scientific, insights into learning. Therefore, just as LA has borrowed insights from learning science up to now, LA has also established a foundation sufficient to provide its own insights to learning science.

% \subsection{学術的インパクト}
% \begin{itemize}
%     \item LAに定義可能性をもたらした: LAとは何をする学問か？ を形式的に定義
%     \item どんなLA実践も公理系の下で統一可能になった
%     \item 理論的研究（概念・哲学）と技術的研究（モデル・手法）の断絶が埋まる → もっとよく議論すべき場所
% \end{itemize}
\subsection{Contribution to LA Itself}
% \begin{itemize}
%     \item 最後に，この理論がLA自身に貢献するところを述べる
%     \item LAは，この理論によって，自身の学術的立場を定義可能になった
%     \begin{itemize}
%         \item 「観測可能性に基づく状態遷移系の科学」としてLAが位置づくことになった
%         \item LAとは何をする学問なのか，LAとは何か，を形式的に定義することができるようになった
%         \item このような定式化によって，発展の方向性が明確になったほか，他分野との対話可能性も向上するであろう
%     \end{itemize}
%     \item また，どのようなLAの実践も公理系の下で統一可能になった
%     \begin{itemize}
%         \item 形式的には異なるデータの取り扱いをしていても，本質的に同じ実践であれば，それらを相互に比較することができるようになった
%         \item （ここに具体例が欲しい！）
%     \end{itemize}
%     \item さらに，LAにおける理論的な研究（概念や哲学）と，技術的な研究（モデル・手法の開発）との断絶を埋められる可能性を示唆した
%     \begin{itemize}
%         \item LAは本来技術主導，工学的関心主導で発展してきた分野だが，そのような実践に理論的基盤が与えられるようになる
%         \item また，理論的考察から新しい技術的発展の可能性が示唆されるようにもなる
%     \end{itemize}
% \end{itemize}
We will discuss how this theory contributes to LA itself. First and foremost, this theory enabled LA to define its own academic standing. Specifically, LA came to be positioned as ``the science of state transition systems based on observability.'' This provided a clear answer to the questions: What does LA study? What can be understood through LA? Through this formulation, LA not only clarified its own developmental direction but also enhanced its potential for dialogue with other fields.

Moreover, in practical research---a symbol of development in LA---it has become possible to treat any case uniformly under a system of axioms. Even when handling data that are formally different, if they are essentially the same practice, they can be compared with each other. This is the first step toward elevating LA to a formal science.

Furthermore, this theory suggests the potential to bridge the gap between theoretical research---such as concepts and philosophy in LA---and technical research focused on developing models and methodologies. While LA has traditionally developed as a technology-driven, engineering-focused field, this approach enables providing a theoretical foundation for such practices. Moreover, theoretical examination can also suggest possibilities for new technical developments.

Regarding ethics, this theory does not directly prescribe what LA should do. However, by mathematically defining the conditions for what can and cannot be called LA, ethical implications naturally arise as a consequence. That is, the fact that this theory's axiomatic system necessitates transparent structuring of observation, experience, state, and inference constitutes its minimal ethical guideline. By excluding machine behaviorism and requiring inferences to be grounded in the learner's context and state, it prevents LA from becoming a form of dangerous data manipulation for learners. This theory ensures that LA does not merely end up with the manipulation of learning/educational data but becomes an endeavor that respects the learner's uniqueness and agency.

\subsection{Scope and Limitation}
% \begin{itemize}
%     \item 公理は「最小十分性」の観点から設計されたが，他の構造もありうること
%     \item 状態の構成は任意性を持ち，モデルに依存すること
%     \item 単一学習者の枠組みであり，学習者同士の協調（協調学習）を考慮できていないこと
%     \begin{itemize}
%         \item しかし，これは学習者それぞれの経験および状態に吸収できるのでは？
%         \item 「観測される行動」の部分が，協調的側面を吸収している
%     \end{itemize}
%     \item 今後，より洗練された公理系が出る可能性（公理系自体の洗練，およびLAの定義自体の変化）
%     \item 社会文化的側面の不在 → 状況的学習論などとの接続は今後の課題
%     \begin{itemize}
%         \item しかし，これも初期状態で吸収できる可能性がある
%     \end{itemize}
% \end{itemize}
% \begin{itemize}
%     \item まず，本公理系は，学習の心理学的定義，LAの要請，最小十分性の観点から設計されたが，他の構造の可能性が排除されたわけではない
%     \begin{itemize}
%         \item 今後より洗練された公理系が出る可能性を排除しない
%         \item また，将来的にLAの定義自体が変わるかもしれない
%         \item このように，未来永劫の価値を持ちうるものであることを保証することはできない
%     \end{itemize}
%     \item 状態の構成が任意性を持ち，モデルに依存する
%     \begin{itemize}
%         \item なるべく簡単な状態空間を設計するという実践的示唆を与えるにとどまる
%     \end{itemize}
%     \item 学習者同士の協調や社会文化的側面を考慮できていない
%     \begin{itemize}
%         \item しかし，これはむしろ本理論の純化のために必要である
%         \item 学習者どうしの協働や，学習者の社会文化的背景は，学習者の初期状態や観測に現れる文脈として，部分的に吸収できる可能性がある
%     \end{itemize}
% \end{itemize}

In this study, we designed an axiomatic system aiming for high generality based on the psychological definition of learning, the requirements of LA, and the principle of minimal sufficiency. Therefore, we deliberately avoided explicitly defining the forms of the experience $e_t \in \mathcal{E}$ or state $s_t \in \mathcal{S}$, allowing designers of LA practices to freely design them. This is not a lack of specificity, but rather an abstraction designed to integrate heterogeneous LA practices and enhance the potential for implementation---an essential design choice. 

This generality and essentiality allow us to state the following. First, the value of this axiom system lies not in determining ``what is LA,'' but rather in determining ``what is not LA.'' That is to say, this axiomatic system, by deliberately broadening the scope of practices that fall under it, avoids unnecessarily narrowing the range of LA. Rather, it enables discussion of the potential for such practices to become LA. Furthermore, design principles for LA practice can be derived from the theorems derived from this axiom system, enabling us to distinguish between ``good'' and ``bad'' LA practice---a practical contribution.

In addition, it is important to note that the concept of LA structure proposed in this study is not limited to LA practices targeting individual learners. For example, the initial state $s_0$ can be defined as a vector containing the states of multiple learners, and other elements can be appropriately defined to target multiple learners as well. This is made possible by the generality of this axiom system, and as a result, this theory can also address the collaborative and social aspects of learning.

% This study designed an axiomatic system based on the psychological definition of learning, the requirements of LA, and the principle of minimal sufficiency.
Although this axiom system was designed with the goals of generality and minimal sufficiency, this does not preclude the possibility of other structures. More refined axiomatic systems may emerge in the future, and the definition of LA itself may also change. Furthermore, stronger axiomatic systems tailored to specific domains may emerge in the future. However, even then, this axiomatic system will likely serve as a useful core.

% This axiomatic system does not account for collaboration among learners or the sociocultural aspects of learning; however, this was necessary to purify the theory. It cannot be ruled out that collaboration among learners and their sociocultural backgrounds could be partially incorporated as context appearing in learners' initial states and observations.

\section{Conclusion and Future Work}
% \begin{itemize}
%     \item LAは応用分野として繁栄してきた
%     \begin{itemize}
%         \item ダッシュボード，成績予測，最適介入の支援
%         \item しかし，そのような成功は，LAという学問のアイデンティティを覆い隠すことにつながっていた
%     \end{itemize}
%     \item 本研究はLAに公理的方法論によって明確な理論的基盤を与える初めての試みである
%     \begin{itemize}
%         \item 学習という営みの心理学的定義，LA実践の要請からスタートした，最小十分な公理である
%         \item これは経験則的に共有されてきたLAの性質を表現可能であり，また任意のLA実践を説明する可能性を持つ
%         \item さらに，この理論によって，LAは自らの立ち位置や哲学的立場を明確にできた
%     \end{itemize}
%     \item Future Work
%     \begin{itemize}
%         \item 経験，状態，評価の具体的な設計原理の探索
%         \item TLAに基づいた新しい「LA実践設計」の理論
%         \begin{itemize}
%             \item 状態の複雑さをなるべく抑えたLA実践の設計とか
%             \item TLAの公理を満たすように設計されたLAシステムの設計とか
%         \end{itemize}
%     \end{itemize}
% \end{itemize}
The field of LA has found application in specific areas of learning support, such as dashboards, performance prediction, and support for optimal interventions. However, such successes have tended to obscure the identity of LA as an academic discipline.

This study represents the first attempt to provide LA with a clear theoretical foundation through an axiomatic methodology. The axiomatic system designed in this study is minimal and sufficient, originating from the psychological definition of the activity of learning and the requirements of LA practice. It can express the properties of LA that have been empirically shared and can provide explanations for any LA practice. Furthermore, this theory has enabled LA to clarify its own academic standing and philosophical position.

The theoretical value of this research lies in elevating LA practice to the subject of mathematical discourse. Through LA structures, analyses previously difficult (e.g., the quality of design, structural constraints, and theoretical limitations) become possible, significantly strengthening the methodological foundation of LA research. Furthermore, the practical value of this research lies in providing design principles for LA practice derived from this axiomatic system. It has become possible to explicitly distinguish between ``good'' and ``bad'' practices---previously shared implicitly---by assessing whether they satisfy the axioms and theorems as requirements.

Future work will explore concrete design principles for experience, state, and inference within this axiomatic system. These elements, which were not given concrete form in this study, warrant more detailed examination. Furthermore, we also seek to construct new theories for designing LA practices based on this framework and explore possibilities for new empirical experiments. Designing LA practices that minimize the complexity of states within the LA structure (e.g., designing $f_s$ that minimizes $|\mathcal{S}|$ while maintaining the necessary expressiveness for capturing learning dynamics, or investigating the properties of $f_e$ that ensure that temporal order information in $O_{\leq t}$ is not lost) would be valuable directions.
Concretely implementing LA systems designed to satisfy the axioms of this theory would also be highly interesting endeavors.

This research has indeed contributed to providing a robust theoretical foundation for LA. As with theoretical contributions in other scientific fields, we conclude this paper with the expectation that this theory of LA will serve as a catalyst for promoting more rigorous, comprehensive, and theory-grounded research.

\section*{Acknowledgment}
This work was supported in part by the Council for Science, 3rd Cross-ministerial Strategic Innovation Promotion Program (SIP),
under Grant JPJ012347;
and in part by Japan Society for the Promotion of Science (JSPS) KAKENHI under Grant 23H00505 and Grant 25K21357.
We also thank Dr. Kento Koike for providing us with useful suggestions.

\bibliography{bibliography}
\bibliographystyle{plain}

\end{document}